\documentclass[conference,a4paper]{IEEEtran}
\IEEEoverridecommandlockouts
\usepackage{cite}
\usepackage{amsmath,amssymb,amsfonts}
\usepackage{algorithmic}
\usepackage{graphicx}
\usepackage{textcomp}
\usepackage{subfigure}
\usepackage{array}
\usepackage{tabularx}
\usepackage{xcolor}
\usepackage{url}
\usepackage{lipsum}
\usepackage{orcidlink}
\usepackage{hyperref}
\def\BibTeX{{\rm B\kern-.05em{\sc i\kern-.025em b}\kern-.08em
    T\kern-.1667em\lower.7ex\hbox{E}\kern-.125emX}}

\begin{document}

\title{\fontsize{21pt}{20pt}\selectfont Generating EUPHEMIA-compatible bids for flexible demand under imperfect information}

\author{
    \IEEEauthorblockN{
        Christian Doh Dinga\orcidlink{0009-0002-1905-8221}{\textsuperscript{a}},
        Mukunda Badarinath{\textsuperscript{b}},
        Seyed Hossein Jamali\orcidlink{0000-0002-4198-0901}{\textsuperscript{c}},
        Laurens de Vries\orcidlink{0000-0002-4014-9294}{\textsuperscript{b}},
        Milos Cvetkovic\orcidlink{0000-0002-5169-7368}{\textsuperscript{a}}
    }
    \\
    \IEEEauthorblockA{
    \textsuperscript{a}\textit{Electrical Sustainable Energy, Delft University of Technology, 2628 CD, Delft, The Netherlands} \\
    \textsuperscript{b}\textit{Engineering Systems and Services, Delft University of Technology, 2600 GA, The Netherlands} \\
    \textsuperscript{c}\textit{Tata Steel Nederland Technology B.V., P.O. Box 10000, 1970 CA IJmuiden, the Netherlands} \\
    \{c.dohdinga, l.j.deVries, m.cvetkovic\}@tudelft.nl, m.badarinath@student.tudelft.nl, seyed.jamali@tatasteeleurope.com
    }
}

\maketitle


\begin{abstract}
Electricity procurement constitutes a significant share of operational costs for large electricity consumers, and thus exposure to extreme prices poses a substantial financial risk. This paper proposes a method to generate EUPHEMIA-compatible bids for flexible demand to enable their participation in the European day-ahead electricity market while minimizing risks. Two strategies are considered, resulting in two bid formats: hourly bids (HBs), representing flexibility via marginal price responsiveness through price–quantity pairs, and exclusive-group bids (EBs), representing flexibility via mutually exclusive operational schedules submitted at opportunity cost. Our method is evaluated on a hypothetical electrolyzer system and a real-world steel plant under different market conditions. Results show that the economic performance of each strategy depends on the operational characteristics of the load and market conditions. Under volatile market conditions, highly flexible systems achieve better economic outcomes with EBs, while less flexible systems with stronger intertemporal constraints perform better with HBs.
\end{abstract}


\section{Introduction}

\begingroup
\renewcommand\thefootnote{}\footnotetext{\vspace{0.5mm}\rule[0.5ex]{0.2\textwidth}{0.5pt}\\979-8-3195-3554-2/26/\$31.00 ©2026 European Union}
\addtocounter{footnote}{-1} 
\endgroup

Electricity procurement accounts for a significant share of operational costs for large electricity consumers, and thus exposure to extreme price spikes poses a substantial financial risk, including increases in production costs that can significantly affect profitability and competitiveness \cite{Karasavvidis2024, HERDING_2024100180}. Many of these consumers have some degree of flexibility\textemdash load shifting\textemdash allowing them to reduce costs by shifting consumption from high to low-price periods. To express this multi-period flexibility during electricity procurement in the European day-ahead electricity market, consumers must submit price-quantity bids compatible with the Pan-European Hybrid Electricity Market Integration Algorithm (EUPHEMIA) \cite{AllNEMO2024} that clears the market for maximum social welfare.

Generating EUPHEMIA-compatible bids for flexible demand is a challenging task for several reasons. For example, consumers need to make decisions regarding their bids for the following day in the face of uncertainty since future electricity prices are not known in advance. Moreover, many flexible loads have operational dependencies that link decisions across time. The bidding algorithm must account for intertemporal constraints, such as how consumption in one hour affects feasible consumption in later hours, across multiple price forecast scenarios to capture uncertainties in market prices.

While existing literature has extensively explored demand-side bidding strategies \cite{HERDING_2024100180, Kardakos_7095586, Zhiwei_7275175, Tianyang_8558521}, a significant knowledge gap remains. Most proposed methods either ignore system operational constraints or generate a single demand time series rather than price–quantity bids. As a result, these approaches cannot fully express consumer flexibility and are not compatible with the bid formats currently supported by EUPHEMIA. Consequently, their practical applicability is limited, as they are inadequate to support consumer participation in the European day-ahead electricity market. An important exception is the work of Karasavvidis \textit{et al.} \cite{Karasavvidis2024}, who propose a method for generating EUPHEMIA-compatible bids in the form of exclusive group bids. However, their approach is demonstrated on a stylized flexible demand system and focuses on a single bid format. As a result, the applicability of the method to real-world industrial consumers with heterogeneous operational constraints and alternative bidding formats remains unclear. 

In this work, we address the following question: \textit{How can flexible industrial consumers generate operationally feasible, EUPHEMIA-compatible bids while managing price uncertainty to optimize their participation in the European day-ahead electricity market?} To answer this question, we develop a method to generate two types of EUPHEMIA-compatible bids: hourly bids (HBs) and exclusive-group bids (EBs). The HBs strategy represents demand flexibility through multiple price–quantity pairs in each hour, forming a stepwise demand curve. This allows consumers to express their willingness to consume electricity at different price levels with high granularity to maximize arbitrage opportunities. The EBs strategy represents flexibility through a set of mutually exclusive consumption profiles. By submitting several alternative consumption schedules, the consumer allows EUPHEMIA to select the profile that maximizes consumer surplus\textemdash and therefore, social welfare\textemdash under the realized market prices. To provide a comprehensive evaluation, these strategies are also compared with two benchmark approaches: a price-insensitive (PI) bidding strategy and a perfect foresight (PF) strategy representing an upper bound on achievable performance. The proposed bidding strategies are evaluated using two case studies. The first case considers a highly flexible electrolyzer system producing green hydrogen, representing a highly price-responsive electricity consumer. The second case considers an existing industrial steel plant in the Netherlands with significant electricity consumption but a more limited flexibility due to operational constraints. For both systems, the bidding strategies are tested under different weather-driven market conditions: a typical weather year with relatively stable electricity prices and a severe weather year characterized by prolonged periods of low renewable generation\textemdash that is, a dunkelflaute\textemdash resulting in higher and more volatile electricity prices. This combination of scenarios allows us to assess how different bidding strategies perform under varying levels of consumer flexibility and price volatility.

In sum, the main contribution of this work is the development of a method for translating multi-period operational flexibility into EUPHEMIA-compatible bids under price uncertainty, enabling flexible electricity consumers to participate more effectively in the European day-ahead electricity market.

The remainder of this paper is structured as follows. Section \uppercase\expandafter{\romannumeral2} presents the methodological framework and mathematical models underlying the two bidding strategies. Section \uppercase\expandafter{\romannumeral3} provides a detailed description of the case studies and experimental setup used to evaluate the strategies. Section \uppercase\expandafter{\romannumeral4} presents and discusses the results. Finally, Section \uppercase\expandafter{\romannumeral5} concludes the paper and outlines directions for future research.


\section{Methodology}
The main goal of this study is to develop a method for translating uncertain day-ahead electricity prices into bids that are operationally feasible and minimize operational costs for the consumer. We first present an overview of the methodological framework and then describe in detail the mathematical formulations underlying the two bidding strategies.

\subsection{Methodology Framework}

\begin{figure*}[htbp]
\centering
\includegraphics[width=\textwidth]{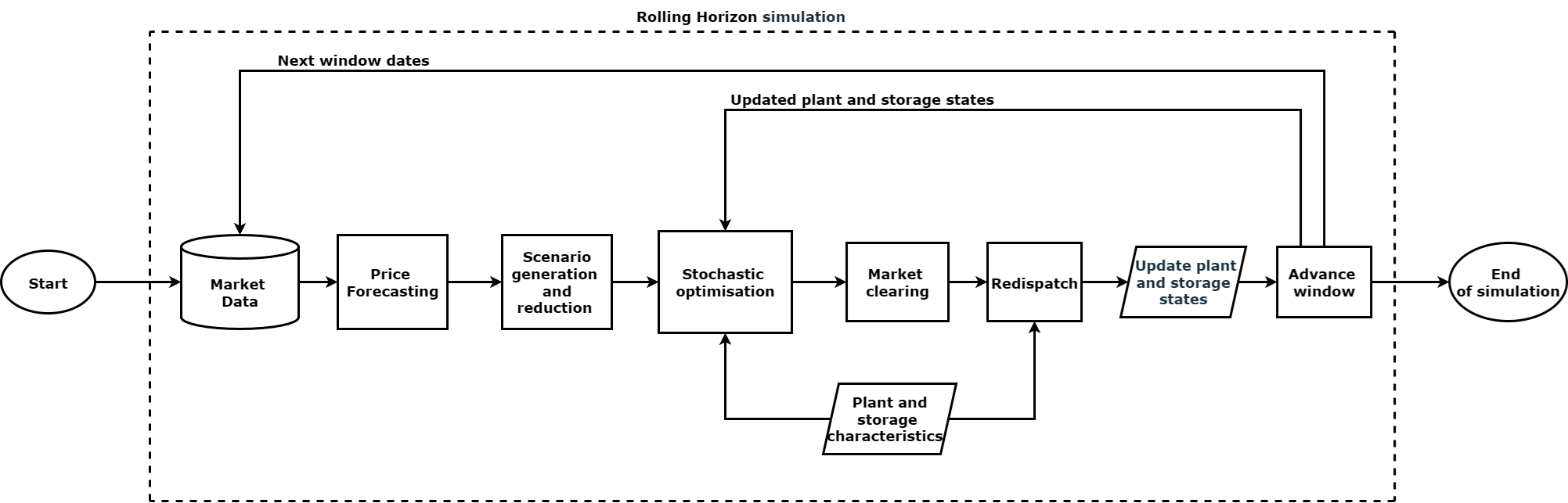}
\caption{Overview of the methodological framework.}
\label{fig:model_framework}
\end{figure*}

The methodological framework is shown in Figure \ref{fig:model_framework}. It consists of four main steps: (1) price forecasting, (2) scenario generation and reduction, (3) stochastic optimization, and (4) rolling-horizon implementation and state updates.



\subsubsection{Price Forecasting}
The first step of the proposed method is to generate a base electricity price forecast, which is then used to construct multiple price scenarios that capture uncertainty in day-ahead market prices. We employ a Prophet-style decomposition model to generate price forecasts. This model is well-suited for electricity price forecasting because electricity prices exhibit multiple seasonal patterns (e.g., daily, weekly, and yearly cycles) and can benefit from the inclusion of exogenous regressors \cite{mishra2024_prophet}. The forecasting model incorporates external predictors such as system load and renewable generation forecasts (wind and solar) to improve predictive performance, consistent with hybrid forecasting approaches commonly used in electricity price forecasting \cite{Huang2024, Bashir2022}. To account for the heavy tails and extreme price spikes frequently observed in electricity markets, the forecast residuals are modeled using a Lévy-stable distribution, enabling generated scenarios to capture rare but significant price events that would not be adequately represented by Gaussian residual models.

\subsubsection{Scenario Generation}

Scenario generation converts the point forecast into a set of plausible price trajectories that represent uncertainty in day-ahead electricity prices. First, random noise is drawn from the fitted Lévy-stable distribution of forecast residuals. Second, realistic intra-day and inter-day temporal correlations are imposed by transforming the initially independent noise vector using the Cholesky factorization of a designed correlation matrix. This step is important because price shocks in electricity markets are typically correlated across hours. Third, the magnitude of uncertainty is scaled with the forecast horizon to reflect the growth of forecasting errors over time, and scenario values are clipped to market price bounds to avoid unrealistic outliers. Finally, to maintain computational tractability of the stochastic optimization problem, the generated scenarios are reduced to a representative subset using K-means clustering, and scenario probabilities are assigned based on the corresponding cluster weights.

\subsubsection{Stochastic Optimization}
While the bid type depends on the bidding strategy, as will be described later, the general bidding problem can be formulated as a two-stage stochastic MILP. The first stage chooses market offers (price-quantity or price-consumption profiles) before prices are revealed. The second stage determines operational dispatch and system feasibility after a price scenario is realized, thereby evaluating the cost (or profit) implications of each first-stage decision.


We control downside risk with Conditional Value-at-Risk (CVaR), combining expected scenario cost with a tail-risk penalty. The risk-averse objective function is:
\begin{equation}
\min \sum_{s \in S} \pi_s \, c_s \;+\; \gamma \left( \zeta \;+\; \frac{1}{1-\alpha}\sum_{s \in S}\pi_s \, \xi_s \right),
\end{equation}
where $c_s$ is the net operational cost in scenario $s$, $\pi_s$ is its probability, $\gamma$ is the risk-aversion weight, $\alpha$ is the CVaR confidence level (e.g., $0.95$), $\zeta$ is the VaR threshold variable, and $\xi_s$ is the excess cost above $\zeta$ in scenario $s$. The tail-loss linking constraints are:
\begin{align}
\xi_s &\ge c_s - \zeta \qquad \forall s \in S,\\
\xi_s &\ge 0 \qquad\qquad\;\;\;\;\; \forall s \in S.
\end{align}


In practice, CVaR may improve rapidly for small $\gamma$ and then plateau. The mechanism is intuitive: once $\gamma$ becomes nonzero, the optimizer has an immediate incentive to eliminate the most expensive tail outcomes (e.g., by reshaping bids to avoid procurement during extreme spike scenarios). After those tail outcomes are already mitigated, further CVaR reduction is limited by (i) hard operational constraints (production targets, ramp limits, storage dynamics), (ii) the limited degrees of freedom imposed by EUPHEMIA-compatible bid formats, and (iii) scenario reduction, which can smooth the tail and reduce marginal gains from increasing $\gamma$.

\subsubsection{Rolling Horizon}


To evaluate performance under realistic sequential decision-making in the day-ahead market, we adopt a rolling-horizon simulation approach. Each day, the optimizer solves a multi-day planning horizon (e.g., five days) to account for look-ahead information such as production targets, storage dynamics, etc. However, only the schedule for the first 24 hours is implemented. After market clearing, the system state (e.g., inventories and unit commitment status) is updated and the horizon rolls forward.

To ensure operational feasibility when the cleared energy profile differs from the optimizer’s preferred schedule, an additional redispatch step is performed using the cleared electricity profile as an input while penalizing unused procured energy. This reflects the practical reality that deviations from the cleared day-ahead schedule can lead to imbalance costs and operational penalties, making feasibility and adherence to the submitted bids or schedule important objectives.

\subsection{Bidding Strategies}

The core of this study is the comparison of two demand-side bidding strategies that are compatible with EUPHEMIA. Both bidding strategies share the same mathematical model but differ in (i) the first-stage decision variables, and (ii) the market-clearing logic, but is in line with how these two different bid formats are cleared in EUPHEMIA \cite{AllNEMO2024}.

For notational brevity, let $t\in T$ denote hours and $s\in S$ price scenarios with probabilities $\pi_s$. Let $\lambda_{t,s}$ be the day-ahead price in hour $t$ under scenario $s$. Let $x_{t,s}$ be the cleared electricity consumption (MWh) passed to the second-stage operational optimization problem, which yields scenario cost $c_s$ used in the risk-averse objective (CVaR). We model market acceptance using standard in-the-money logic consistent with each bid structure. Note that the full EUPHEMIA clearing algorithm is not replicated because we only consider a single consumer and therefore is assumed to be a price-taker.

\subsubsection{Hourly Bids (HBs)}

Hourly bids treat each hour as a separate clearing opportunity. The model submits a stepwise demand curve by offering quantities $q_{t,b}$ at bid prices $\lambda^{\mathrm{bid}}_{t,b}$ for each hour $t$ and bid block $b\in B$, with a non-increasing willingness-to-pay structure $\lambda^{\mathrm{bid}}_{t,1}\ge \cdots \ge \lambda^{\mathrm{bid}}_{t,|B|}$.

First-stage variables are the offered quantities $q_{t,b}\ge 0$. Scenario-dependent acceptance is represented by $y_{t,b,s}\in\{0,1\}$, and cleared consumption is
\begin{align}
x_{t,s} &= \sum_{b\in B} q_{t,b}\, y_{t,b,s} && \forall t,s. \label{eq:hourly_x}
\end{align}
The in-the-money logic is enforced via
\begin{align}
\lambda_{t,s} &\le \lambda^{\mathrm{bid}}_{t,b} + M(1-y_{t,b,s}) && \forall t,b,s, \label{eq:hourly_itm}
\end{align}
and step consistency (higher-price blocks cannot be rejected while lower-price blocks are accepted) via
\begin{align}
y_{t,b,s} &\ge y_{t,b+1,s} && \forall t,\, b<|B|,\, s. \label{eq:hourly_monotone}
\end{align}
where $M$ is a sufficiently large constant (big-$M$) based on known price bounds, e.g., $M=\lambda^{\max}-\lambda^{\min}$, used to relax in-the-money constraints when a bid/profile is not accepted.
Equations \eqref{eq:hourly_x}--\eqref{eq:hourly_monotone} define how hourly bids can partially clear hour-by-hour and thus create inter-temporal discontinuities in procurement.

Hourly bids create an agile risk profile: they can selectively avoid high-price hours, reducing exposure to spikes. However, partial clearing can create hour-to-hour discontinuities in procurement. If an hour clears at a low quantity during a price spike, the system may be forced to compensate in neighboring hours to meet production and balance constraints (ramping limits, minimum stable consumption levels, and storage dynamics). This can cause aggressive ramping or even infeasibility if the system is more constrained.

\subsubsection{Exclusive-group Bids (EBs)}

Exclusive-group bids constitute a set of mutually exclusive consumption profiles submitted are a single price (e.g., opportunity cost). Each profile $k\in K$ specifies a coherent quantity vector $Q_{t,k}\ge 0$ over hours $t$ and a profile bid price $\Lambda^{\mathrm{bid}}_{k}$. Compared to hourly bids, acceptance implies a full-day (or multi-hour) schedule that is operationally feasible due to a coherent construction.

First-stage profile selection uses $z_k\in\{0,1\}$:
\begin{align}
\sum_{k\in K} z_k &\le 1. \label{eq:excl_one}
\end{align}
Scenario-dependent acceptance is represented by $a_{k,s}\in\{0,1\}$ and linked to selection by
\begin{align}
a_{k,s} &\le z_k && \forall k,s. \label{eq:excl_link}
\end{align}
Cleared consumption is all-or-nothing for the accepted profile:
\begin{align}
x_{t,s} &= \sum_{k\in K} Q_{t,k}\, a_{k,s} && \forall t,s. \label{eq:excl_x}
\end{align}
To impose price-dependent clearing in a compact way, we enforce an in-the-money condition based on a weighted-average price over the profile hours:
\begin{align}
\bar{\lambda}_{k,s} := \sum_{t \in T} w_{t,k}\lambda_{t,s}, \qquad
\bar{\lambda}_{k,s} &\le \Lambda^{\mathrm{bid}}_{k} + M(1-a_{k,s}) && \forall k,s, \label{eq:excl_itm}
\end{align}
where $w_{t,k}$ are fixed weights satisfying $\sum_{t\in T} w_{t,k}=1$ for each $k$ (e.g., chosen proportional to $Q_{t,k}$ over the hours covered by profile $k$). Constraints \eqref{eq:excl_one}--\eqref{eq:excl_x} enforce mutual exclusivity and eliminate partial intra-day clearing, thereby reducing inter-temporal recovery costs and feasibility risk.

The EB strategy generally exhibits a more conservative risk profile, reducing the likelihood of partial clearing and the associated redispatch costs. However, the trade-off is reduced flexibility to avoid isolated high-price hours, since the bid is submitted as a bundled schedule rather than through hour-by-hour price responsiveness or price arbitrage. Consequently, exclusive-group bidding tends to yield lower variance and more robust feasibility, but may result in higher average costs when price spikes occur over short durations.

In both strategies, first-stage bidding variables determine $x_{t,s}$ through the clearing constraints, and $x_{t,s}$ feeds the second-stage operational MILP to compute scenario cost $c_s$, which is optimized under the CVaR-based risk-averse objective.

\section{Case Study Description}

The proposed bidding framework is evaluated on two representative industrial electricity demand cases: (A) a flexible electrolyzer and (B) a steel production site reflecting an electricity procurement structure used in practice. In both cases, bids are submitted to the day-ahead market for a 24-hour horizon within a rolling-horizon simulation framework. The same price forecasting, scenario generation, and CVaR-based risk formulation are applied in both case studies for a more consistent comparison between the bidding strategies.

\subsection{Electrolyzer}

The first case considers a hypothetical grid-connected water electrolyzer producing hydrogen. The unit operates between minimum and maximum power limits of 5.5 MW and 55 MW, respectively, and is subject to ramp-rate constraints of 52.25 MW/h. Hydrogen production is coupled with a storage system with a capacity of 10 ton, enabling intertemporal load shifting while meeting production targets. The electrolyzer has a production rate of 18 kg/MWh and can adjust electricity consumption in response to price signals within these operational limits. Due to its high ramping capability and relatively weak intertemporal coupling, this system represents a highly flexible industrial load and allows the bidding strategies to actively exploit price variability or arbitrage opportunities.

\subsection{Steel Plant}

The second case represents a steel production site reflecting an electricity procurement structure used in current industrial practice of Tata Steel in the Netherlands. Electricity demand is driven by tightly coupled production processes, ranging from blast furnaces to hot strip mills, which impose binding minimum load requirements, limited short-term flexibility, and strong intertemporal dependencies. Compared to the electrolyzer, the system is therefore subject to significantly stricter operational constraints. Maintaining feasibility requires coordinated multi-hour operating schedules, making more coherent bid profiles particularly important in market participation.

\section{Results}

\subsection{Bids}
\begin{figure*}[htbp]
    \hspace*{10mm}{\large\bfseries Appendix}  
    \par\vspace{8mm}
    \centering
    \subfigure[Electrolyzer: hourly bids in normal weather]{
    \includegraphics[height=0.17\textheight,trim={0 0 0 0},clip]{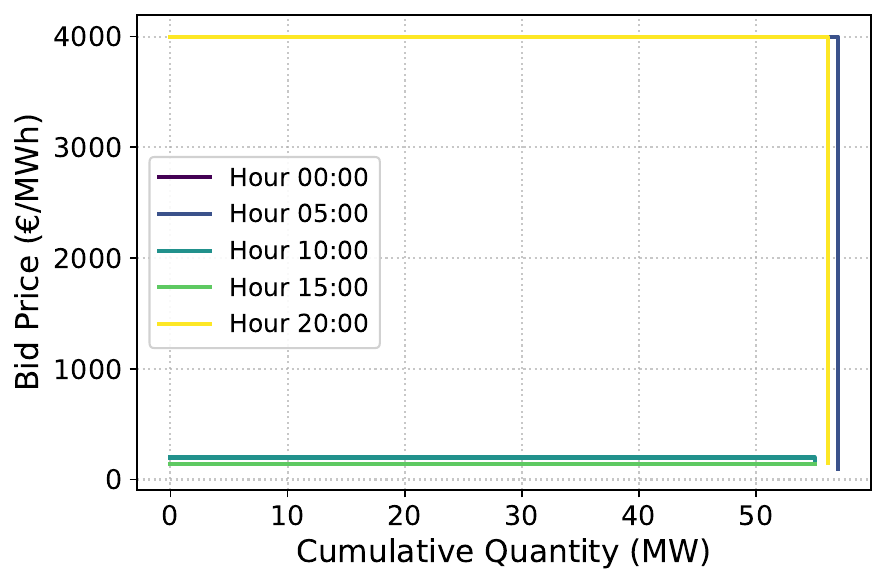}}
    \subfigure[Electrolyzer: hourly bids in Dunkelflaute]{
    \includegraphics[height=0.17\textheight,trim={0 0 0 0},clip]{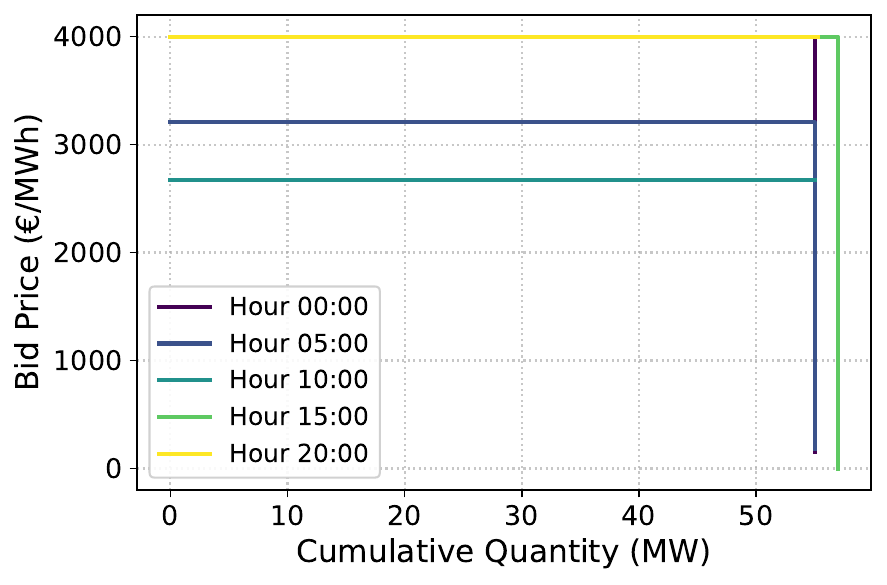}}

    \par\medskip

    \subfigure[Steel plant: hourly bids in normal weather]{
    \includegraphics[height=0.17\textheight,trim={0 0 0 0},clip]{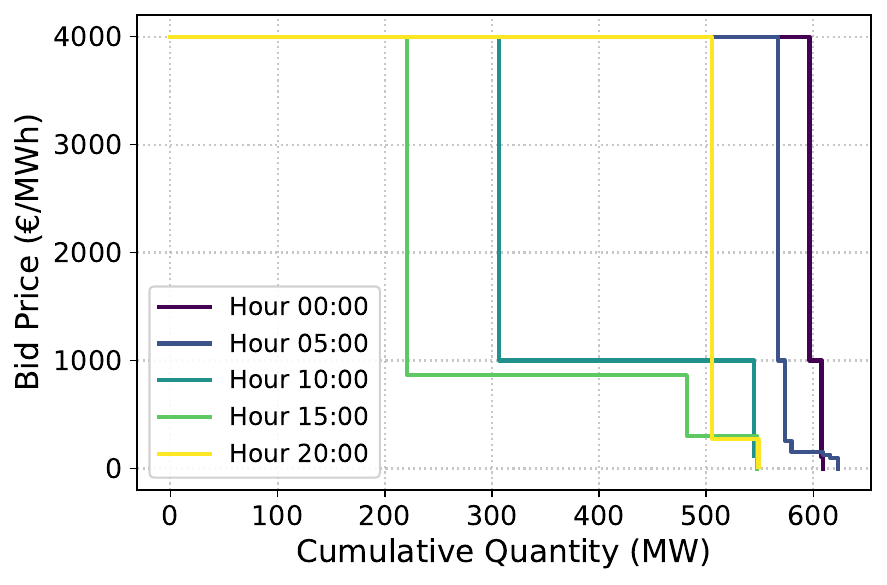}}
    \subfigure[Steel plant: hourly bids in Dunkelflaute]{
    \includegraphics[height=0.17\textheight,trim={0 0 0 0},clip]{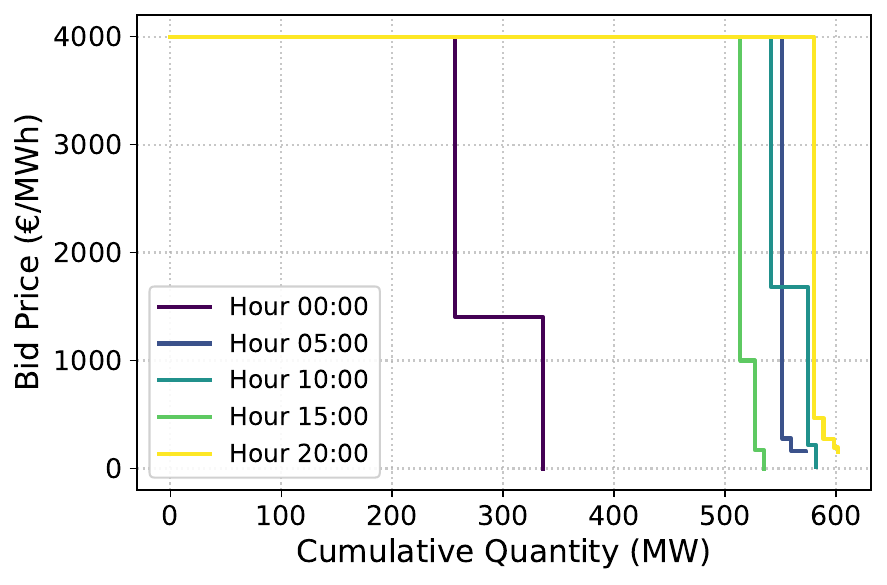}}

    \caption{Hourly bids for the two case studies under different market conditions}
    \label{fig:hourly_bids}
\end{figure*}

\begin{figure*}[htbp]
    \vspace{3mm}
    \centering
    \subfigure[Electrolyzer: exclusive-group bids in normal weather]{
    \includegraphics[height=0.17\textheight,trim={0 1.6cm 0 0},clip]{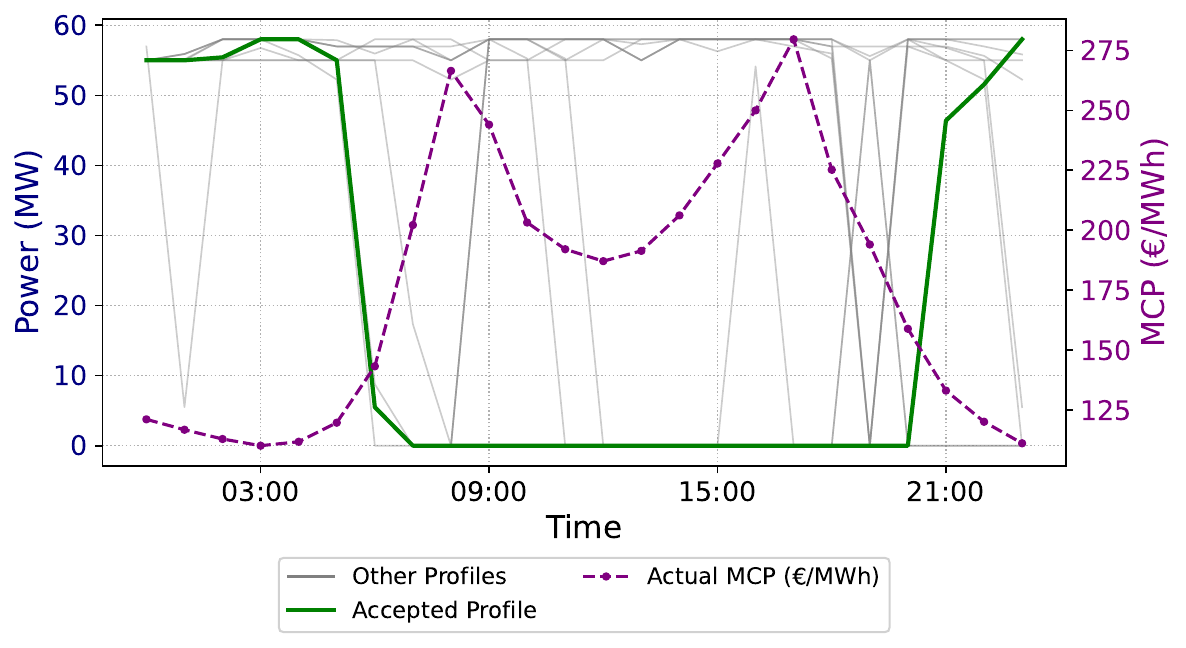}}
    \subfigure[Electrolyzer: exclusive-group bids in Dunkelflaute]{
    \includegraphics[height=0.17\textheight,trim={0 1.6cm 0 0},clip]{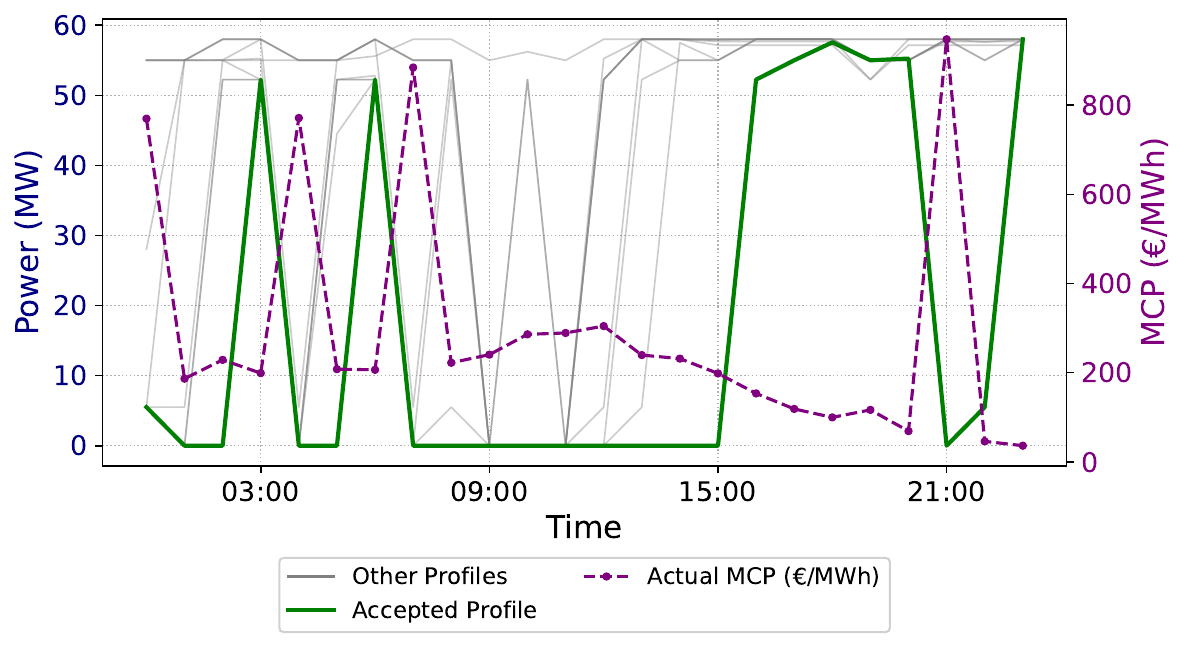}}

    \par\medskip

    \subfigure[Steel plant: exclusive-group bids in normal weather]{
    \includegraphics[height=0.2\textheight,trim={0 0 0 0},clip]{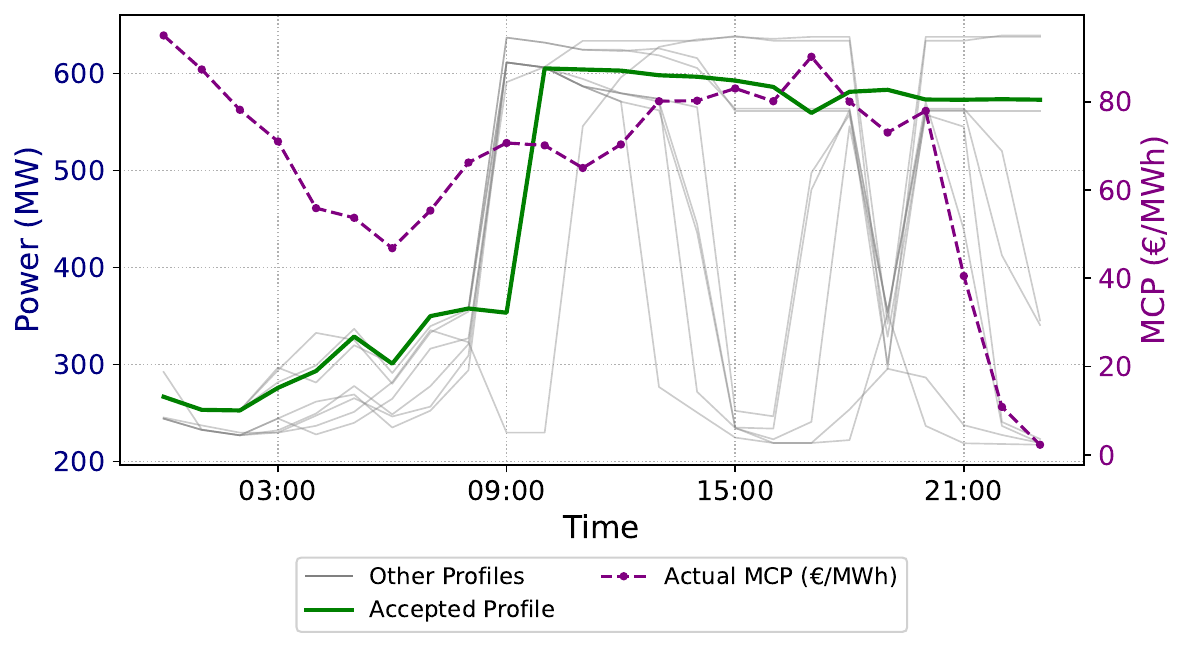}}
    \subfigure[Steel plant: exclusive-group bids in Dunkelflaute]{
    \includegraphics[height=0.2\textheight,trim={0 0 0 0},clip]{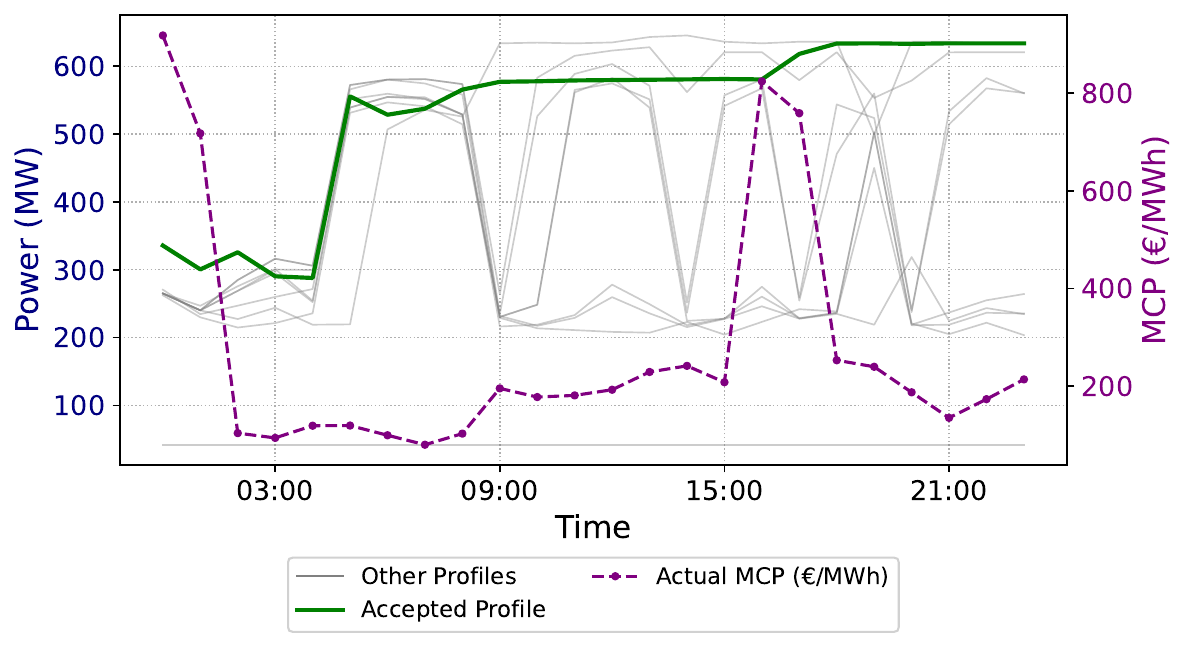}}

    \caption{Exclusive-group bid for the two case studies under different market conditions (MCP: Market-Clearing Price)}
    \label{fig:exclusive_group_bids}
\end{figure*}

Figures~\ref{fig:hourly_bids} and~\ref{fig:exclusive_group_bids} show the HBs and EBs for the two case studies under different scenarios. The HBs in Figure~\ref{fig:hourly_bids} reveal clear differences in bid structure between the electrolyzer and the steel plant. For the electrolyzer system, HBs generally contain fewer price–quantity steps. Because the electrolyzer can ramp almost instantaneously with limited intertemporal coupling, market-clearing outcomes in one hour do not strongly constrain operation in the next. The bidding strategy can therefore concentrate demand into one or a few large bid steps, effectively representing a threshold-based decision to operate when prices fall below a certain level.

In contrast, the steel plant faces stricter ramping limits and stronger intertemporal dependencies. To avoid infeasible ramp transitions when bids are accepted in one hour but not the next, the HBs strategy distributes demand across more price levels, resulting in bid curves with more steps. Under the severe weather scenario with higher and more volatile prices, this effect becomes more pronounced for the steel plant, as additional bid steps help hedge against intertemporal feasibility risks while remaining responsive to price signals. The key finding is that the number of steps in hourly bids is not a direct measure of flexibility, but rather reflects how finely the system must hedge against intertemporal feasibility risks arising from uncertainty.

Figure~\ref{fig:exclusive_group_bids} shows the EBs. The gray lines represent submitted profiles—operationally feasible schedules generated under different price scenarios—while the green line shows the accepted profile selected by EUPHEMIA. For the electrolyzer, the submitted profiles exhibit substantial diversity, reflecting its high flexibility and ability to shift production across hours. The accepted profile therefore tends to concentrate consumption in the lowest-price periods, particularly during the more volatile severe weather scenario. In contrast, the steel plant exhibits a narrower range of feasible profiles due to ramping limits and stronger intertemporal coupling, resulting in more clustered schedules and smaller adjustments in the accepted profile.

Overall, these results highlight a key difference between the two bid formats. While HBs express marginal price responsiveness in each hour, EBs allow the market to choose between entire operational schedules, enabling flexible consumers to communicate intertemporally coherent schedules that ensure operational feasibility.

\subsection{Electricity costs}
\begin{figure*}[htbp]
    \vspace{-50mm}
    \centering
    \subfigure[Electrolyzer]{
    \includegraphics[height=0.2\textheight,trim={0 0cm 0cm 0},clip]{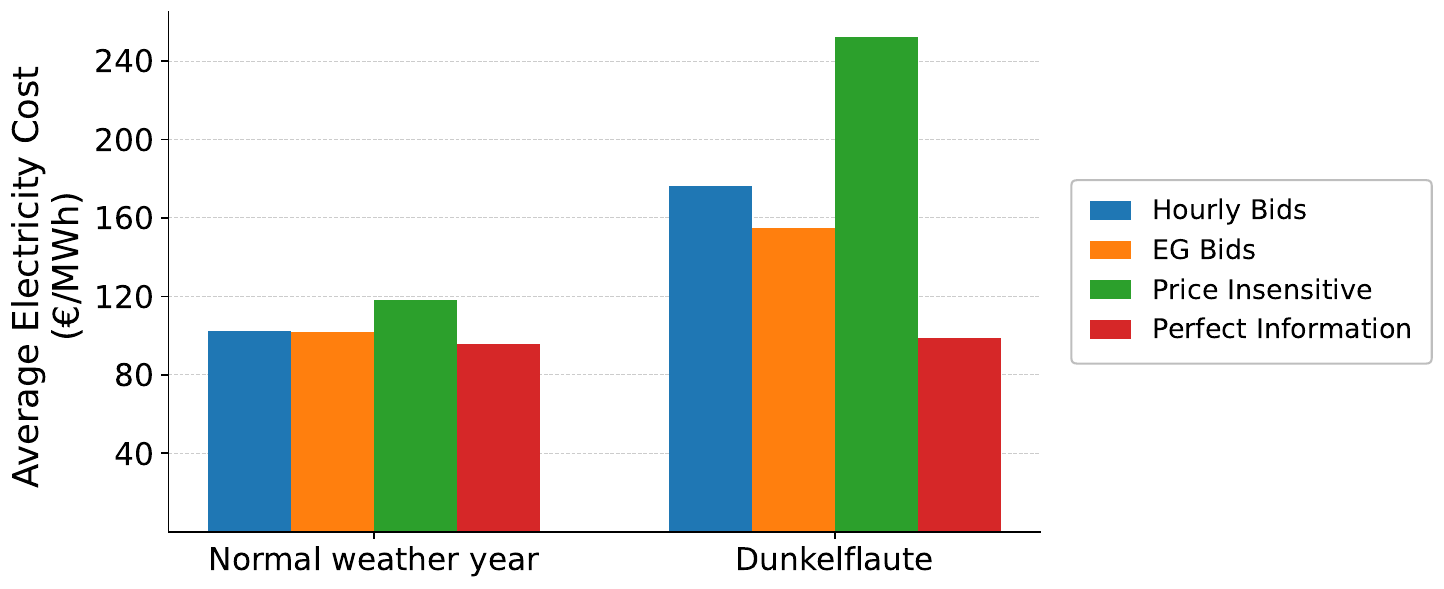}}
    \hspace{0.1em} 
    \par\medskip
    \subfigure[Steel plant]{
    \includegraphics[height=0.2\textheight,trim={0 0 0cm 0},clip]{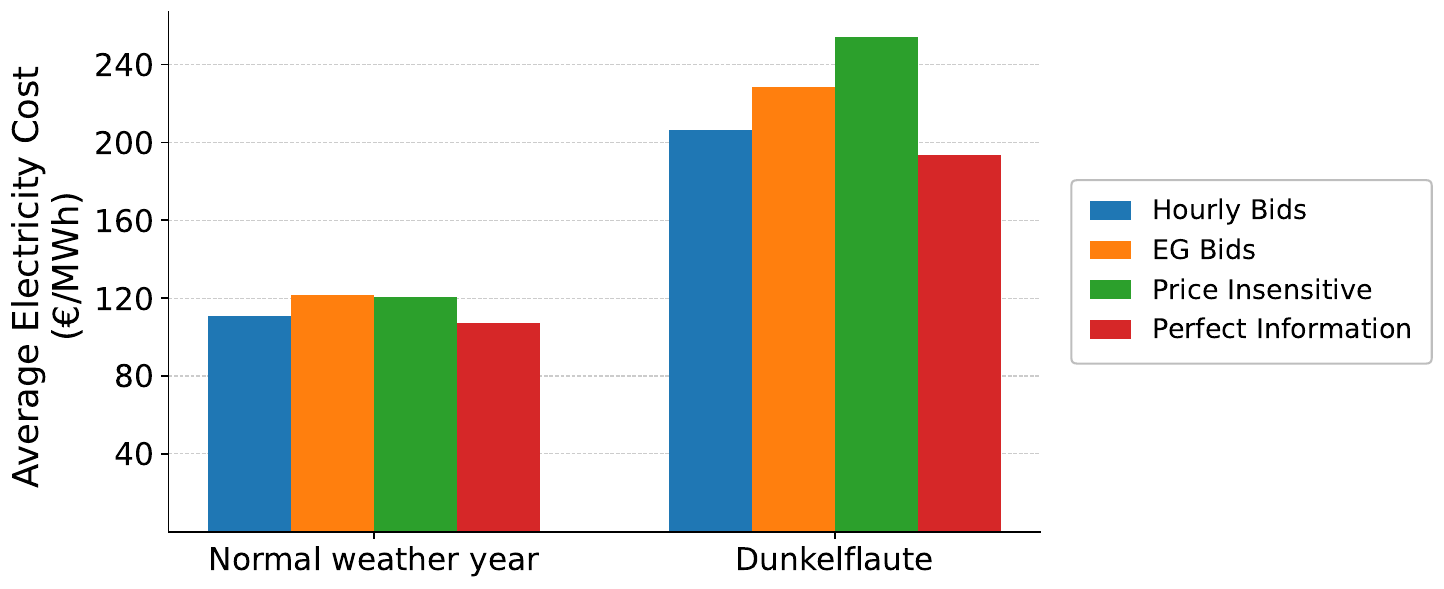}}

    \caption{Average electricity costs for different bidding strategies under different market conditions}
    \label{fig:electricity_costs}
\end{figure*}

Figure~\ref{fig:electricity_costs} compares the average electricity costs obtained under the different bidding strategies for both case studies and weather scenarios. The average electricity cost is defined as the total electricity expenditure divided by the total electricity consumption over the considered period. For the electrolyzer system, HBs generally result in lower costs than EBs and price-insensitive bids under the normal weather scenario and perform close to the perfect foresight benchmark. Under the severe weather scenario with higher and more volatile prices, however, EBs perform better and achieve results closest to perfect foresight, as the multiple submitted profiles allows the market to select the profile that best matches the realized price trajectory. For the steel plant, the HBs strategy consistently achieves lower costs than the EBs strategy irrespective of the weather scenario. This reflects the more constrained operational characteristics of the steel production process, where ramping limits and intertemporal dependencies restrict the range of feasible operating schedules that can make more freely ramp up and down for price arbitrage.

Overall, these results suggest that the economic performance of each bidding strategy depends on the level of operational flexibility and market conditions: loads with higher flexibility can benefit from exclusive group bids under volatile prices, while load with lesser flexibility and strong intertemporal dependencies perform better with more granular hourly bids.

\section{Conclusion}
This paper presented a method for generating operationally feasible, EUPHEMIA-compatible bids for flexible electricity consumers under price uncertainty. Two bidding strategies were proposed: hourly bids (HBs), which express marginal price responsiveness through price–quantity pairs in each hour, and exclusive group bids (EBs), which represent flexibility through alternative operational schedules.

The results show that the two bid formats capture flexibility in fundamentally different ways. While HBs allow consumers to respond to prices on an hourly basis, EBs allow the market-clearing algorithm to select among complete operational schedules that satisfy intertemporal constraints. Consequently, the economic performance of each strategy depends on the operational characteristics of the load and prevailing market conditions. Highly flexible systems, such as electrolyzers, can benefit from EBs under volatile price conditions, whereas systems with stronger intertemporal dependencies, such as the steel plant, tend to perform better with granular hourly bids.

More broadly, the need for such bidding algorithms reflects a structural issue in the current market design: without short-term forward contracts, consumers must make decisions about their future electricity consumption under incomplete information. This work, therefore, demonstrates how electricity consumers with different levels of flexibility can translate uncertain price information into operationally feasible EUPHEMIA-compatible bids to facilitate their participation in the European day-ahead electricity market.

\section{Acknowledgments}

This publication is part of the project DEMOSES (with project number ESI.2019.004 of the research programme Energy System Integration which is (partly) financed by the Dutch Research Council (NWO). 

\bibliographystyle{IEEEtran}
\bibliography{references}

\newpage

\setlength{\floatsep}{1pt}

\setlength{\floatsep}{1pt}

\end{document}